\documentclass[aps, prd, twocolumn, amsmath, floats,floatfix, superscriptaddress, nofootinbib,
showpacs]{revtex4-1}
\usepackage{enumerate}
\usepackage{graphicx}
\usepackage{dcolumn}
\usepackage{bm}
\usepackage{subfigure}

\begin{document}
\title{On the instability for massive scalar fields in Kerr-Newman spacetime}
\author{Yang Huang}
\email{saisehuang@163.com}
\author{Dao-Jun Liu}%
\email{djliu@shnu.edu.cn}
\author{Xiang-hua Zhai}%
\email{zhaixh@shnu.edu.cn}
\author{Xin-zhou Li}%
\email{kychz@shnu.edu.cn}
\affiliation{Center for Astrophysics and Department of Physics, Shanghai Normal University, 100 Guilin Road, Shanghai 200234, China}%
\date{\today}

\begin{abstract}
It is known that a massive charged scalar field can trigger a superradiant instability in the background of a Kerr-Newman black hole. 
In this paper, we present a numerical study of such an instability by using the continued fraction method. It is shown that for given a black hole, 
the unstable scalar mode with a specific azimuthal index $m$ only occurs in a finite region in the parameter space of the scalar field.
The maximum mass of the scalar cloud is exactly the upper bound of the mass of the unstable modes. 
We show that due to the electromagnetic interaction between the scalar field and the Kerr-Newman black hole, 
the growth rate of the instability can be $15.7\%$ larger than that of a scalar field in Kerr spacetime of the same rotation parameter. 
In addition, we find a maximum value of the growth rate $\tau^{-1}=1.788\times10^{-7}M^{-1}$, which is about $4\%$ larger than that in the Kerr case.  

\end{abstract}

\maketitle

\section{Introduction}

In their pioneering work, Regge and Wheeler showed that a Schwarzschild black hole remains stable
under a small nonspherical perturbation \cite{Regge:1957td}. 
Since then, the issue of black hole stability has played a crucial role in general relativity \cite{Zerilli:1970se,Vishveshwara:1970cc,Zerilli:1971wd,Teukolsky:1972my,Chandrasekhar1983,Konoplya:2011qq}. 
It is well established that a Kerr-Newman back hole, the most general asymptotically flat solution in the Einstein-Maxwell theory \cite{Newman:1965my},
is stable under (non)-linear massless perturbations \cite{Zilhao:2014wqa,Dias:2015wqa}.
However, the Kerr-Newman black hole may be unstable due to the "black hole bomb" effect, 
proposed by Press and Teukolsky in \cite{Press:1972zz}, in which the black hole mass and charge could be 
extracted by a charged scalar field through the superradiance mechanism \cite{Damour:1976kh}. Consider a scalar 
wave of the form $e^{-i\omega t-im\phi}$ incident on a Kerr-Newman black hole of mass $M$, electric charge $Q$, 
and angular momentum $J=aM$. If the frequency satisfies the superradiant condition \cite{Bekenstein:1973mi}
\begin{equation}\label{Eq: the superradiant condition}
	\omega<\omega_c\equiv\frac{ma+qQr_+}{r^2_++a^2},
\end{equation}
where $q$ is the electric charge of the scalar field, then the scattered wave is 
amplified. According to the law of conservation of energy, the extra energy of this process comes from the black hole. 
If the scalar field  has a nonzero mass, then it can form bound states in the vicinity of the black hole. 
Bound states in the superradiant regime (\ref{Eq: the superradiant condition}) can extract the rotational energy from the black hole repeatedly, 
triggering a superradiant instability in the background. For a recent review of superradiance, see Ref.~\cite{Brito:2015oca}.

The superradiant instability of massive scalar fields on the Kerr(-Newman) black hole background 
has been studied in many contexts. It was shown more than forty years ago that the Kerr-Newman black hole is unstable 
under a massive charged scalar perturbation \cite{Damour:1976kh}. 
The growth rate of the unstable modes for massive (neutral) scalar fields around a Kerr black hole has been studied in the 
regime $\mu M\gg1$ and $\mu M\ll1$, where $\mu$ is the mass of the scalar field \cite{Zouros:1979iw,Detweiler:1980uk}. 
In Ref.~\cite{Furuhashi:2004jk}, Furuhashi and Nambu computed the bound state frequencies analytically in the regime $\mu M\ll1$, $qQ\ll1$, 
and numerically in the regime $\omega\sim\mu$. Then, Dolan studied the superradiant instability for a massive (neutral) scalar 
field around a Kerr black hole for arbitrary values of $\mu M$, using the continued fraction method \cite{Dolan:2007mj}.
Besides, many authors have studied the extremal black hole bomb analytically \cite{Rosa:2009ei,Hod:2009cp}.

Recently, some of us studied the charged massive scalar clouds around a Kerr-Newman black hole \cite{Huang:2016qnk}. 
It was pointed out that for a given value of $m$, the necessary conditions for the superradiant instability
are fulfilled only in a finite region in the parameter space of the scalar field.
One of the motivations of this paper is to confirm the prediction that the instability only occurs in the finite region.
To this end, we compute the bound state spectrum in the whole unstable region. 
Furthermore, this paper also aims to compute the accurate upper limits for the growth rate of the unstable modes. 
In the neutral case \cite{Dolan:2012yt}, Dolan found a maximum growth rate for $a=0.997M$, $l=m=1$, and $\mu M=0.45$, with $\tau^{-1}\equiv\omega_I=1.72\times10^{-7}M^{-1}$. 
At this point, it is natural to ask: (i) whether the instability could be enhanced by introducing the electromagnetic interaction
between the scalar field and the black hole? (ii) If enhanced, then what is the maximum growth rate? To the best of 
our knowledge, these issues have not been addressed.

The paper is organized as follows. In Sec.~\ref{Sec: background} we describe the basic equations and boundary conditions
of the eigenvalue problem for the bound states of a massive charged scalar field in Kerr-Newman spacetime. 
Next, in Sec.~\ref{Sec: potential analysis}, we derive the necessary conditions for the bound states and the superradiant 
instability. In Sec.~\ref{Sec: Numerical approach}, we describe the continued fraction method to compute the bound state 
spectra. Then, the main results are presented in Sec.~\ref{Sec: result}. 
Finally, we conclude in Sec.~\ref{Sec: conclusion}. Throughout the paper, we use the units in which $G=\hbar=c=1$.

\section{Scalar field in Kerr-Newman spacetime}\label{Sec: background}

We consider a massive charged scalar field, $\Psi$, with mass $\mu$ and charge $q$, propagating in the background of
a Kerr-Newman black hole. In Boyer-Lindquist coordinates, the line element of the background spacetime reads
\begin{equation}\label{Eq: line element}
	\begin{aligned}
	ds^2=&-\frac{\Delta}{\rho^2}\left(dt-a\sin^2\theta d\phi\right)^2+\frac{\rho^2}{\Delta}dr^2+\rho^2d\theta^2
	\\&+\frac{\sin^2\theta}{\rho^2}\left[\left(r^2+a^2\right)d\phi-adt\right]^2,
	\end{aligned}
\end{equation}
with $\rho^2\equiv r^2+a^2\cos^2\theta$, and $\Delta\equiv r^2-2Mr+a^2+Q^2$. The Kerr-Newman 
black hole has two horizons:
\begin{equation}
	r_\pm=M\pm\sqrt{M^2-a^2-Q^2}.
\end{equation}
When the rotation parameter $a$ is fixed, the maximum charge of the Kerr-Newman black hole is given by 
\begin{equation}
	Q_{\text{max}}\equiv\sqrt{M^2-a^2}.
\end{equation}
When $Q=Q_{\text{max}}$, the black hole becomes extremal and we have $r_+=r_-$. The background electromagnetic 
potential is 
\begin{equation}
A_{\alpha}=\left(-\frac{Qr}{\rho^2},0,0,\frac{aQr\sin^2\theta}{\rho^2}\right).
\end{equation}

The propagation of a massive charged scalar field in curved spacetime is described by the Klein-Gordon equation
\begin{equation}\label{Eq: Klein-Gordon eq}
	\left(\nabla_\alpha-iqA_\alpha\right)\left(\nabla^\alpha-iqA^\alpha\right)\Psi=\mu^2\Psi.
\end{equation}
Substituting the decomposition
\begin{equation}
	\Psi=\sum_{lm}R_{lm}(r)S_{lm}(\theta)e^{im\phi}e^{-i\omega t},
\end{equation}
into Eq.~(\ref{Eq: Klein-Gordon eq}), we obtain the angular equation
\begin{equation}\label{Eq: Angular eq}
	\begin{aligned}
	\frac{1}{\sin\theta}&\frac{d}{d\theta}\left(\sin\theta\frac{dS_{lm}}{d\theta}\right)
	\\&+\left[A_{lm}+a^2(\omega^2-\mu^2)\cos^2\theta-\frac{m^2}{\sin^2\theta}\right]S_{lm}=0,
	\end{aligned}
\end{equation}
and the radial equation
\begin{equation}\label{Eq: Radial eq}
	\Delta\frac{d}{dr}\left(\Delta\frac{dR_{lm}}{dr}\right)+UR_{lm}=0,
\end{equation}
where
\begin{equation}
	\begin{aligned}
	U=&\left[\omega\left(r^2+a^2\right)-am-qQr\right]^2
	\\&+\Delta\left[2am\omega-A_{lm}-\left(a^2\omega^2+\mu^2r^2\right)\right].
	\end{aligned}
\end{equation}
Here, $A_{lm}$ is the separation constant. If $a^2(\mu^2-\omega^2)\lesssim m^2$, then $A_{lm}$ may be expanded as 
a power series
\begin{equation}
	A_{lm}=a^2(\mu^2-\omega^2)+\sum_{k=0}^{\infty}c_ka^{2k}(\mu^2-\omega^2)^k.
\end{equation}
where $c_0=l(l+1)$ and other coefficients $c_k$ may be found in \cite{Olver:2010:NHMF,NIST:DLMF}. Beyond this regime, 
one can compute the angular eigenvalue $A_{lm}$ numerically with the continued-fraction method \cite{Leaver:1985ax}. 

Near the event horizon $r=r_+$, the radial function behaves as
\begin{equation}\label{Eq: asymptotic behavior at the horizon}
	R_{lm}(r\rightarrow r_+)\sim e^{\pm i(\omega-\omega_c)r_*},
\end{equation}
where the minus sign describes the ingoing wave at the horizon, which is the physically accepted solution, 
and the tortoise coordinate $r_*$ is defined by
\begin{equation}
	\frac{dr_*}{dr}\equiv\frac{r^2+a^2}{\Delta}.
\end{equation}

At spatial infinity, the asymptotic solution of the radial equation (\ref{Eq: Radial eq}) reads
\begin{equation}
	R_{lm}(r\rightarrow\infty)\sim r^{\chi-1}e^{k r},
\end{equation}
where
\begin{equation}
	\chi=\frac{M(\mu^2-2\omega^2)+qQ\omega}{k},
\end{equation}
and $k=\pm\sqrt{\mu^2-\omega^2}$. The choice of sign for $k$ determines the behavior at infinity. 
The well-known quasinormal modes (QNMs) correspond to $\mathrm{Re}(k)>0$, which describes purely outgoing waves at infinity. 
In this paper, we are interested in the bound state solutions, which decay exponentially at infinity ($\mathrm{Re}(k)<0$).

The radial equation (\ref{Eq: Radial eq}), together with the boundary conditions discussed above, 
becomes an eigenvalue problem for $\omega$ and only a discrete set of complex frequencies 
(expressed by $\omega=\omega_R+i\omega_I$) is allowed. 
The imaginary part $\omega_I$ sets the growth ($\omega_I>0$) or decay ($\omega_I<0$) rate of the amplitude of the scalar field.

\section{The regime of superradiant instabilities}\label{Sec: potential analysis}

The existence of bound states (and possibly, the unstable modes) is guaranteed by the binding potential well outside the black hole. 
In our previous work, we analyzed the effective potential of the radial function, and found that the necessary condition for 
the bound state is (see \cite{Huang:2016qnk} for details),
\begin{equation}\label{Ineq: 0}
	f(\mu,q)<\omega<\mu,
\end{equation}
where
\begin{equation}\label{Def: f(u,q)}
	f(\mu,q)\equiv\frac{qQ}{4M}+\sqrt{\frac{\mu^2}{2}+\frac{q^2Q^2}{16M^2}}.
\end{equation}
Bearing in mind the superradiance condition $\omega<\omega_c$, one obtains the necessary conditions for the superradiant instability:
\begin{equation}\label{Ineq: I}
	f(\mu,q)<\omega<\omega_c<\mu,
\end{equation}
or
\begin{equation}\label{Ineq: II}
	f(\mu,q)<\omega<\mu<\omega_c.
\end{equation}

Note that, when $m\leq0$, the two inequalities can not be satisfied, which implies that there is no unstable mode in this case. 
For a positive $m$, inequalities (\ref{Ineq: I}) and (\ref{Ineq: II}) form a closed region in the parameter space of the scalar 
field\footnote{A plane spanned by the mass $\mu$ and charge $q$ of the scalar field. We also refer to it as the $(\mu,q)$ plane in this paper.}. 
The superradiant instability is expected to appear in this region.

\section{The eigenvalue problem for quasibound states}\label{Sec: Numerical approach}
 
In this paper, we apply the continued fraction method \cite{Leaver:1985ax} to compute the bound state frequencies.
Solutions of Eq.(\ref{Eq: Radial eq}) satisfying the ingoing wave condition at the horizon and decaying exponentially at infinity 
admit the following series expansion
\begin{equation}\label{Eq: series expansion}
	R_{lm}(r)=\left(r-r_-\right)^{\chi-1}e^{kr}\sum_{n=0}^{\infty}a_{n}\left(\frac{r-r_+}{r-r_-}\right)^{n-i\sigma},
\end{equation}
where
\begin{equation}
	\sigma=\frac{\left(r^2_++a^2\right)\left(\omega-\omega_c\right)}{r_+-r_-}.
\end{equation}

Substituting Eq.(\ref{Eq: series expansion}) into Eq.(\ref{Eq: Radial eq}) yields a three-term recurrence relation for the coefficients $a_n$,
\begin{equation}\label{Eq: the ratio of a_0}
	\alpha_0a_1+\beta_0a_0=0,
\end{equation}
\begin{equation}
	\alpha_n a_{n+1}+\beta_n a_n+\gamma_n a_{n-1}=0,\;\;n>0,
\end{equation}
where
\begin{equation}
	\alpha_n=n^2-(c_0-1)n-c_0,
\end{equation}
\begin{equation}
	\beta_n=-2n^2-2c_1n-c_2,
\end{equation}
and
\begin{equation}
	\gamma_n=n^2-c_3n-c_4.
\end{equation}
Coefficients $c_1,c_2,c_3$ and $c_4$ are given by
\begin{widetext}
	\begin{equation}\label{eq: c0}
		c_0=-1+2i\omega-iqQ+\frac{2i}{b}\left(\omega-\frac{am+qQ+\omega Q^2}{2}\right),
	\end{equation}
	\begin{equation}
	c_1=-c_0-(1+2b)k+\frac{\omega(\omega-qQ)}{k},
	\end{equation}
	\begin{equation}
		c_2=A_{lm}+(b+1)^2 k^2+\left(-2b+Q^2-2\right)\omega^2+2\omega((b+1)qQ+i)-iqQ-(c_1+c_0+1+iqQ-2i\omega)c_0,
	\end{equation}
	\begin{equation}
		c_3=1+c_0+\frac{2\left(\mu^2+qQ\omega-2\omega^2\right)}{k},
	\end{equation}
	and
	\begin{equation}\label{eq: c4}
		c_4=(k+i\omega)(k+i\omega-iqQ)\left(\frac{2i\omega-iqQ}{k}-\frac{1+c_0+c_3}{2k}\right).
	\end{equation}
\end{widetext}
Here, $b\equiv\sqrt{1-a^2-Q^2}$. Note that in Eqs.~(\ref{eq: c0})-(\ref{eq: c4}), we take $M$ as scale such that all physical quantities are dimensionless in these equations.
It is easy to check that when $Q=0$, our expressions for $\alpha_n,\beta_n$, and $\gamma_n$ reduce to those in Ref.~\cite{Dolan:2007mj}. 
Then, the ratio of successive $a_n$ is given by an infinite continued-fraction
\begin{equation}\label{Eq: the ratio of a_n}
	\frac{a_{n+1}}{a_n}=\frac{-\gamma_{n+1}}{\beta_{n+1}-}\frac{-\gamma_{n+2}}{\beta_{n+2}-}\frac{-\gamma_{n+3}}{\beta_{n+3}-}\cdots.
\end{equation}
Evaluating (\ref{Eq: the ratio of a_n}) at $n=0$ and substituting into Eq.(\ref{Eq: the ratio of a_0}), we obtain the characteristic equation for the bound state frequency
\begin{equation}
	0=\beta_0-\frac{\alpha_0\gamma_1}{\beta_1-}\frac{\alpha_1\gamma_2}{\beta_2-}\frac{\alpha_2\gamma_3}{\beta_3-}\cdots.
\end{equation}
Then the bound state frequency is obtained by minimizing the absolute value of the right side of this equation.

\section{Results}\label{Sec: result}

We have computed the quasibound state frequencies numerically via the continued fraction method, and the main results 
are summarized in Figs.~\ref{fig: most unstable neutral}-\ref{fig: omegaVSq}. In these plots, we exhibit the properties 
of the quasibound state spectrum in the parameter space of the scalar field. 
Here, we are interested in the most unstable modes. Hence, we only display the results for $l=m=1$ and $n=0$, where $n$ is the excitation number.

\subsection{Neutral scalar field}

\begin{figure*}
	\includegraphics[width=0.32\textwidth,height=0.2\textheight]{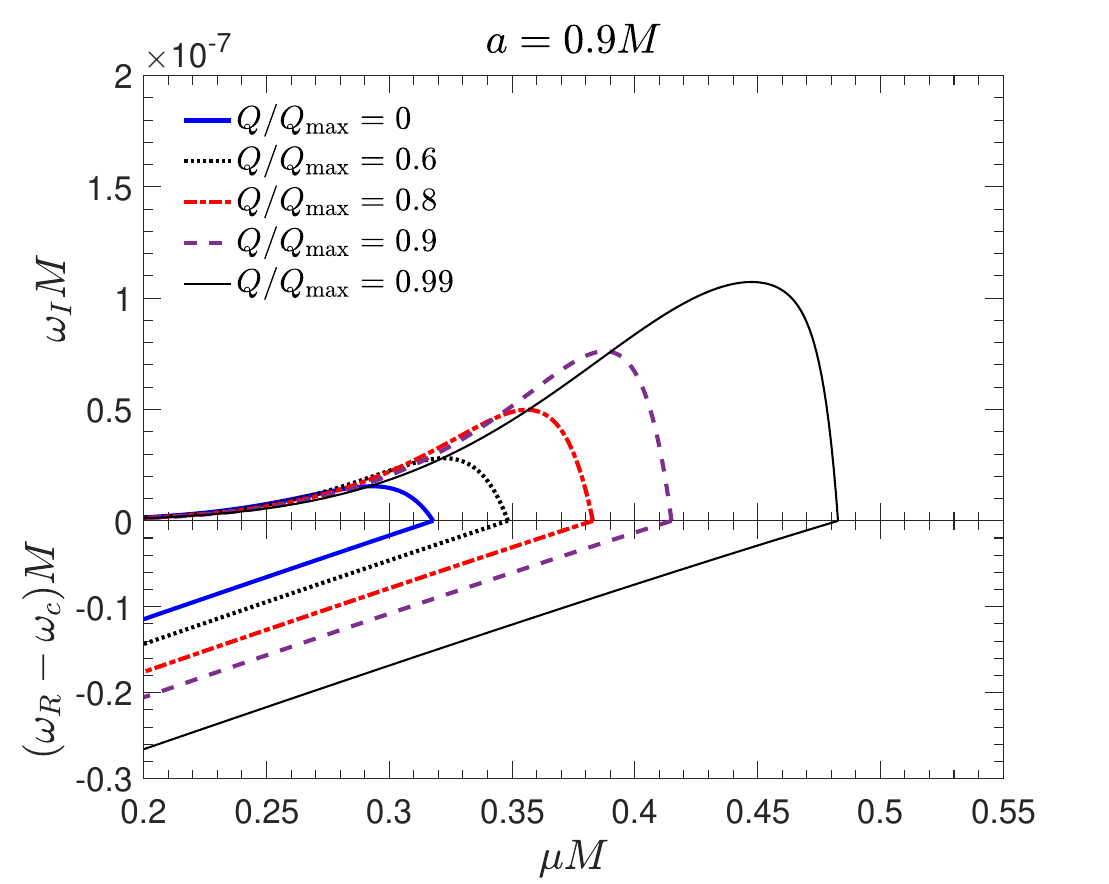}	
	\includegraphics[width=0.32\textwidth,height=0.2\textheight]{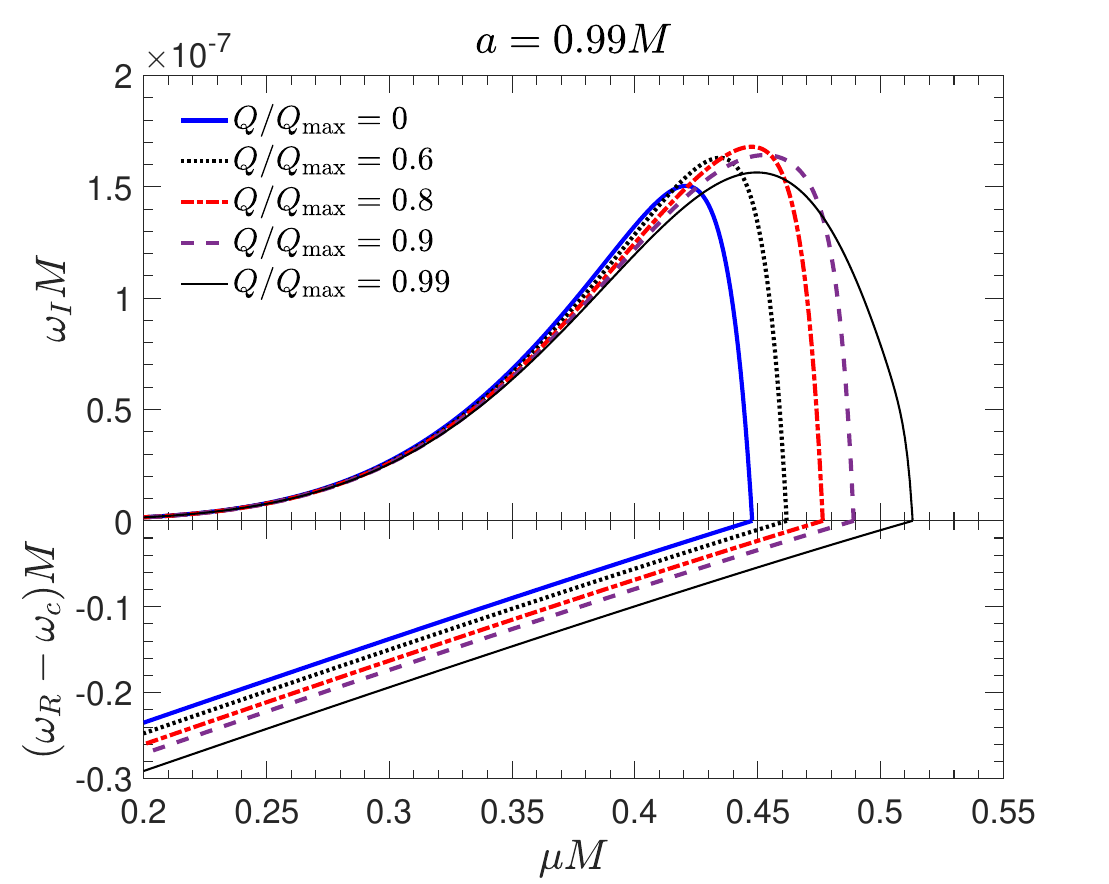}	
	\includegraphics[width=0.32\textwidth,height=0.2\textheight]{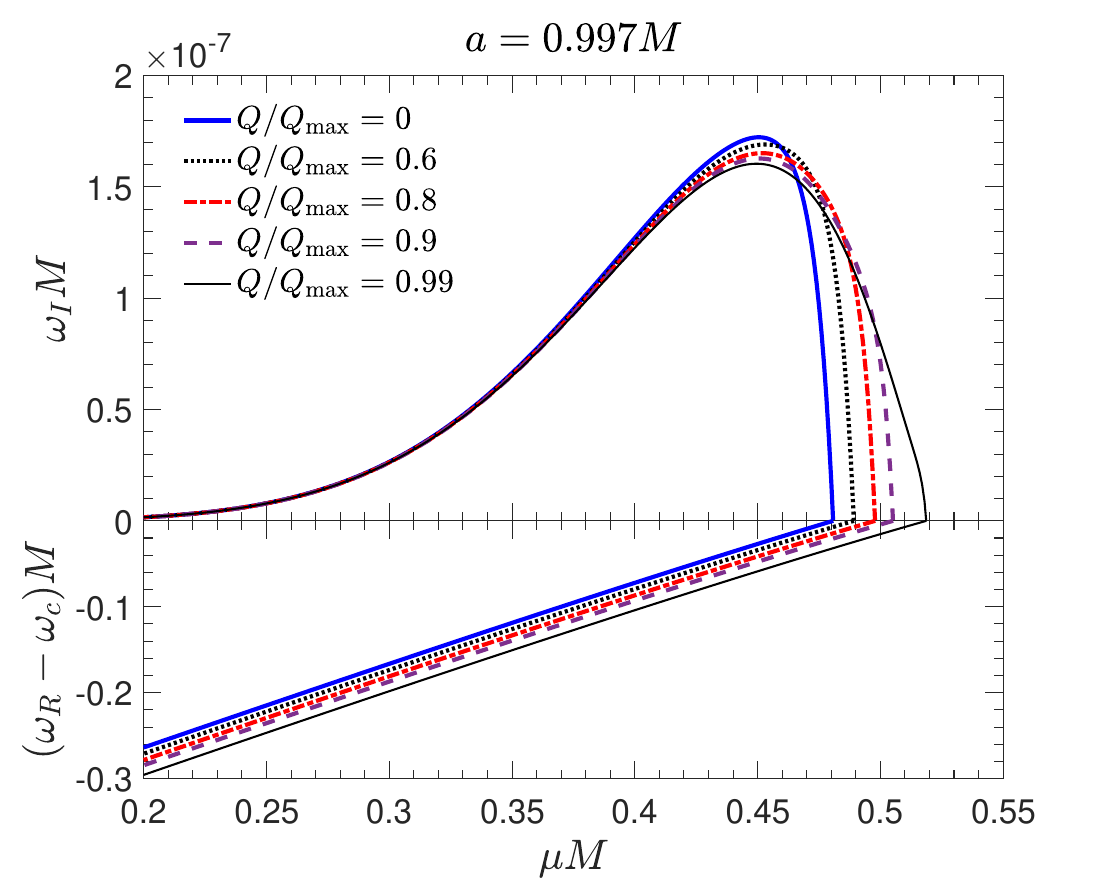}
	\caption{Bound state spectrum of a neutral scalar field ($l=m=1$) for different values of $a$ and $Q$. 
	The rotation parameters of the black hole are $a=0.9,0.99,0.997M$, from left to right. The plot shows clearly that
	the instability $\omega_I>0$ occurs when the superradiant condition $0<\omega_R<\omega_c$ is satisfied.}
	\label{fig: most unstable neutral}
\end{figure*}

A massive scalar field in Kerr-Newman spacetime has a rich spectrum. We first consider a neutral scalar field. 
Figure.~\ref{fig: most unstable neutral} shows the effects of $Q$ on the quasibound state ($l=m=1$) frequency, 
where the rotation parameter of the black hole is fixed as $a=0.9,0.99,0.997M$ from left to right. Firstly, the plot shows 
clearly that the instability ($\omega_I>0$) always happens within the superradiant regime, $0<\omega_R<\omega_c$.
Second, for given values of $a$ and $Q$, the growth rate $\omega_I$ peaks at some $\mu M$. In general, the peak value of
$\omega_I$ does not vary monotonously with $Q$. For example, when $a=0.99M$ we find a peak value of 
$\omega_IM\approx1.68\times10^{-7}$ for $Q/Q_{\text{max}}\approx0.8$ and $\mu M\approx0.448$. However, if
$a\gtrsim0.997$, the peak value of $\omega_I$ is decreased as $Q$ grows. Hence, for a neutral scalar field 
in the Kerr-Newman background, the greatest growth rate occurs at $a=0.997M$, $Q=0$, and $\mu M=0.45$, with
$\omega_IM=1.72\times10^{-7}$, which is in consistent with the result in Ref.~\cite{Dolan:2012yt}.

\subsection{Charged scalar field}\label{subSec: unstable region}

As discussed in Sec.~\ref{Sec: potential analysis}, for a charged scalar field, the superradiant instability
only occurs in the region determined by conditions (\ref{Ineq: I})-(\ref{Ineq: II}). This is confirmed in Fig.~\ref{fig: ContourPlot}, 
where the mass and charge dependence of $M\omega_I$ are displayed. Two auxiliary reference lines are also plotted for making up the boundary of this region:

\begin{figure}
	\includegraphics[width=0.5\textwidth,height=0.25\textheight]{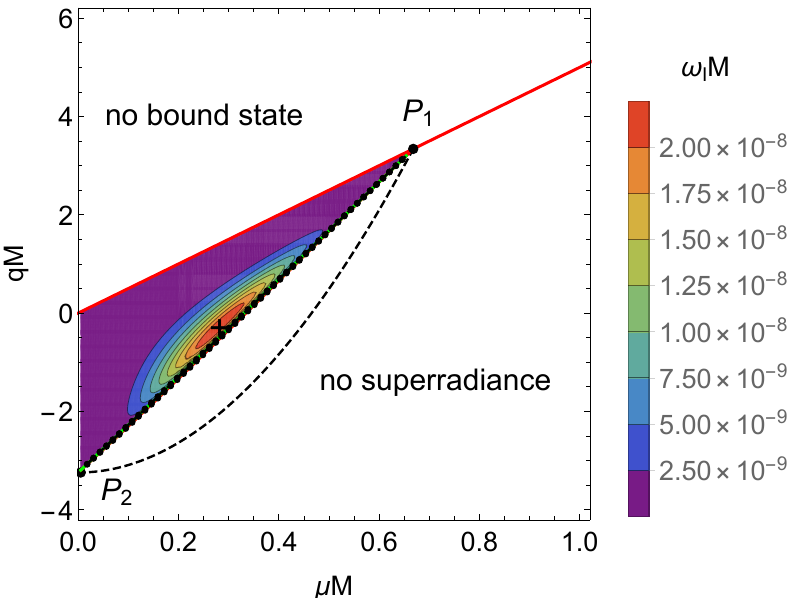}
	\caption{The dependence of the mass $\mu$ and charge $q$ on the imaginary part $\omega_I$ of the quasibound state frequency, 
	with $Q=0.2M$, $a=0.9M$, and $l=m=1$. The plot confirms that superradiant instability only happens in the region determined 
	by inequalities (\ref{Ineq: I})-(\ref{Ineq: II}). The plus symbol represents the location of the most unstable mode, i.e., 
	$(\mu M,qM)\approx(0.282,-0.264)$, with $\omega_I M\approx2.243\times10^{-8}$.}
	\label{fig: ContourPlot}
\end{figure}

\begin{enumerate}[(a)]
	\item\label{line: a} $f(\mu,q)=\mu$ (red solid). Using the definition of $f(\mu,q)$ in Eq.(\ref{Def: f(u,q)}), we find that 
	this line simply gives
	\begin{equation}
	\mu M=qQ.
	\end{equation}
	This is the threshold line for quasibound states. Above this line, no bound state can be found since the bound state condition 
	(\ref{Ineq: 0}) can not be satisfied in this case. Note that in Ref.~\cite{Furuhashi:2004jk}, the authors found a condition 
	for the bound state to exist, i.e., $\mu M>qQ$. According to their analysis, this condition only works in the small coupling parameters approximation, i.e., $|qQ|\ll1$ and $\mu M\ll1$. 
	However, from inequality (\ref{Ineq: 0}), we conclude that this condition is also valid for large coupling parameters.
	
	\item\label{line: b} $f(\mu,q)=\omega_c$ (black dashed). This is a branch of hyperbolic curve in the $(\mu,q)$ plane
	\begin{equation}
	qQ=\frac{AB+\sqrt{A^2+8(B^2-1)M^2\mu^2}}{B^2-1},
	\end{equation}
	where
	\begin{equation}
	A=4mM\Omega_H,\;\;\;B=1-\frac{4Mr_+}{r^2_++a^2}.
	\end{equation}
	Below this line we have $\omega_c<f(\mu,q)$, which implies that no unstable modes could be found in this case, 
	since the quasibound state condition (\ref{Ineq: 0}) is not compatible with the superradiance condition.
\end{enumerate}

\begin{figure*}
	\includegraphics[width=0.43\textwidth,height=0.2\textheight]{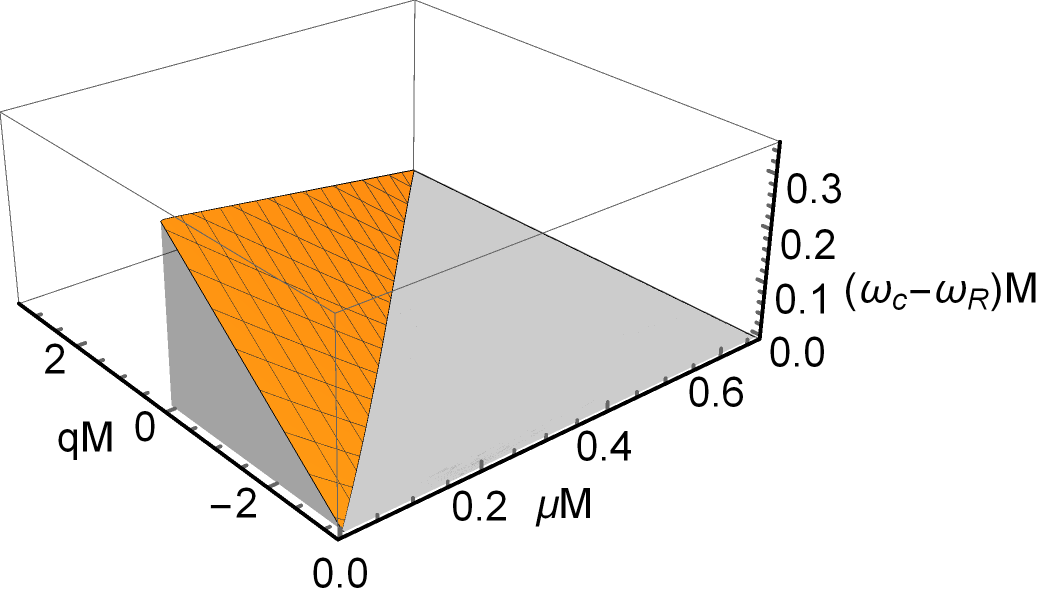} \label{fig: 3D Real part}
	\includegraphics[width=0.42\textwidth,height=0.2\textheight]{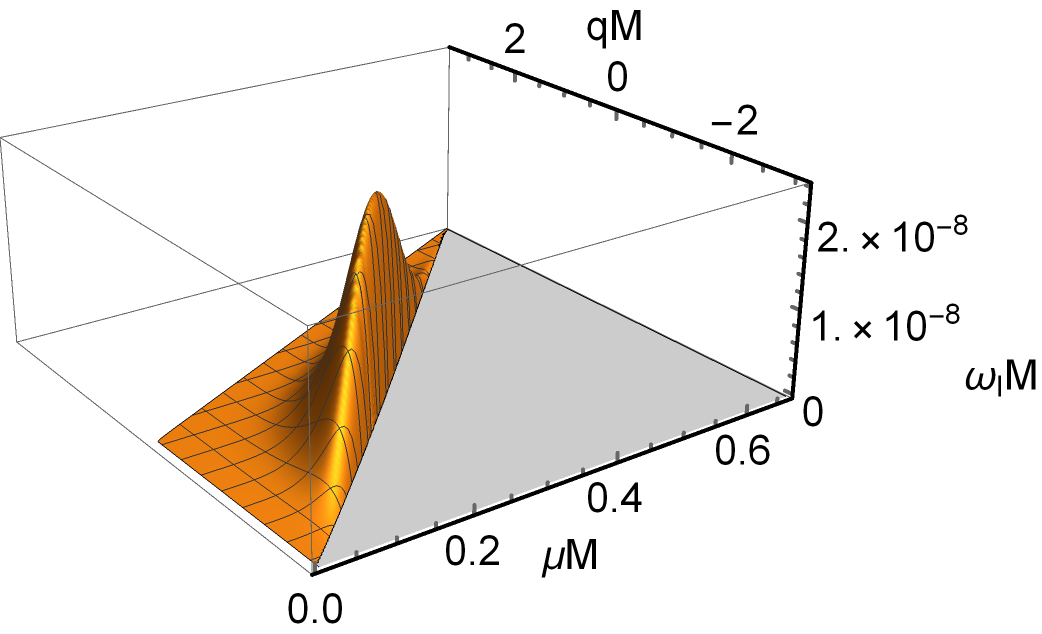} \label{fig: 3D Imaginary part}
	\caption{
		Frequency spectrum of the bound state in the unstable region. Parameters in this plot are the same as those in Fig.~\ref{fig: ContourPlot}. 
		Left panel: the dependence of the dimensionless mass $\mu M$ and charge $qM$ on $(\omega_c-\omega_R)M$. 
		Right panel: the imaginary part $\omega_I$ as a function of $\mu M$ and $qM$. This plot shows inside the unstable region, 
		the superradiance condition $\omega_R<\omega_c$ is satisfied, and $\omega_I>0$. 
		}
	\label{fig: 3D plot}
\end{figure*}

Figure.~\ref{fig: ContourPlot} shows that, as expected, the superradiant instability happens only in the region between lines (\ref{line: a}) and (\ref{line: b}).
Note that inequalities (\ref{Ineq: I})-(\ref{Ineq: II}) are merely necessary conditions. 
Thus, the unstable region\footnote{By "unstable region" we mean the region in which the superradiant instability ($\omega_I>0$) happens.} 
does not fully cover the region surrounded by lines (\ref{line: a}) and (\ref{line: b}). The exact onset curve 
(the dotted line in Fig.~\ref{fig: ContourPlot}) of the superradiant instability can be obtained from the numerical method used in \cite{Huang:2016qnk,Benone:2014ssa}. 
Each point on the onset curve corresponds to a stationary state, with $\omega_R=\omega_c$ and $\omega_I=0$. 
This is the so-called scalar clouds \cite{Benone:2014ssa,Huang:2016qnk,Hod:2012px,Herdeiro:2014goa,Hod:2016yxg,Hod:2014baa,Bernard:2016wqo,Huang:2017whw}. 
As is shown in \cite{Huang:2016qnk}, the onset curve has two endpoints, i.e, $P_1$ and $P_2$ in the plot, 
which determine the mass (and charge) range of the scalar cloud. It is found that $P_1$ is the intersection of lines (\ref{line: a}) and (\ref{line: b}), 
and $P_2$ is the intersection of line (\ref{line: b}) and the $q$-axis. By definition, coordinates of points $P_1$ and $P_2$ are respectively given by
\begin{equation}\label{Eq: upper bound mass}
(\mu_1,q_1)=\left(\frac{ma}{Mr_+-Q^2},\frac{mMa}{Q(Mr_+-Q^2)}\right),
\end{equation}  
and
\begin{equation}
(\mu_2,q_2)=\left(0,-\frac{ma}{Qr_+}\right).
\end{equation}
From Fig.~\ref{fig: ContourPlot}, we see that the onset curve and line (\ref{line: a}) make up the boundary of the unstable region. 
Hence, points $P_1$ and $P_2$ also determine the upper and lower bound of the mass (charge) of the unstable modes, respectively. 
To validate that the instability we have found is indeed caused by superradiance, we present the full spectrum in the unstable region in Fig.~\ref{fig: 3D plot}. 
The plot shows clearly that the superradiance condition $\omega_R<\omega_c$ is always satisfied in the unstable region.

\begin{figure*}
	\includegraphics[width=0.45\textwidth,height=0.27\textheight]{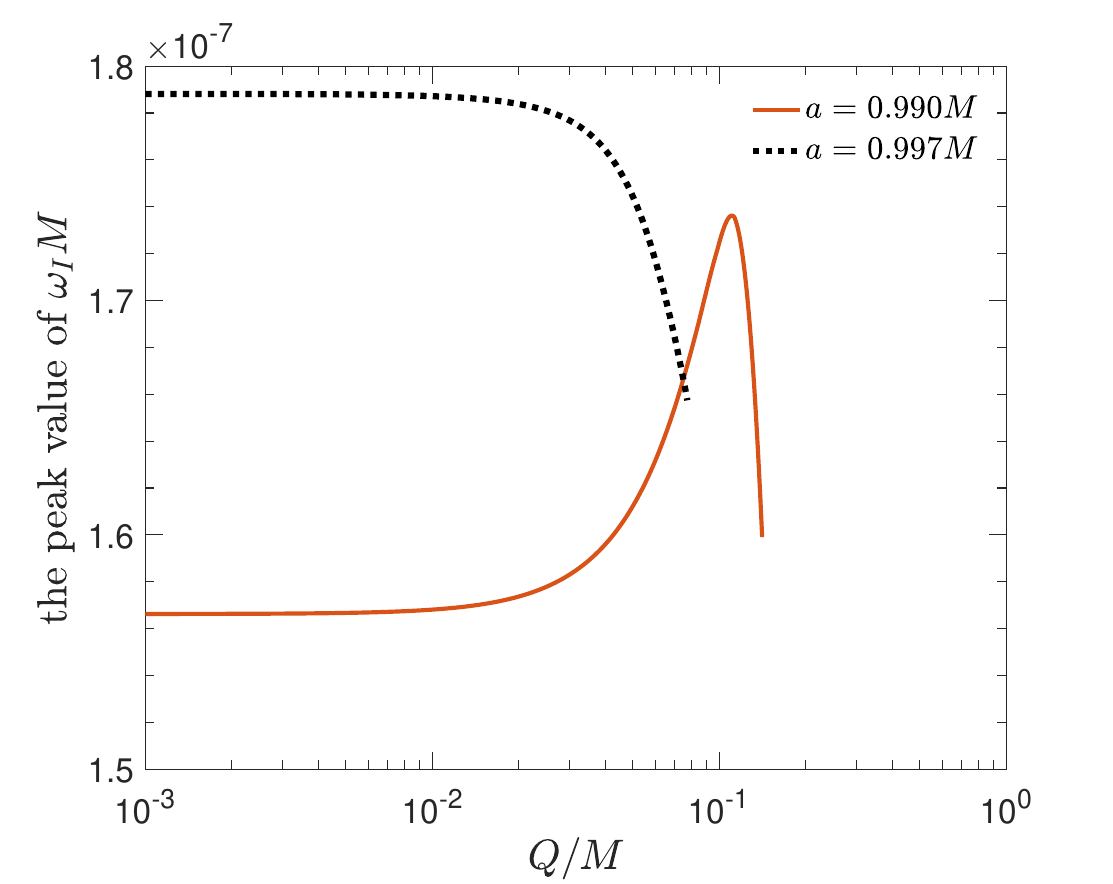}
	\includegraphics[width=0.45\textwidth,height=0.27\textheight]{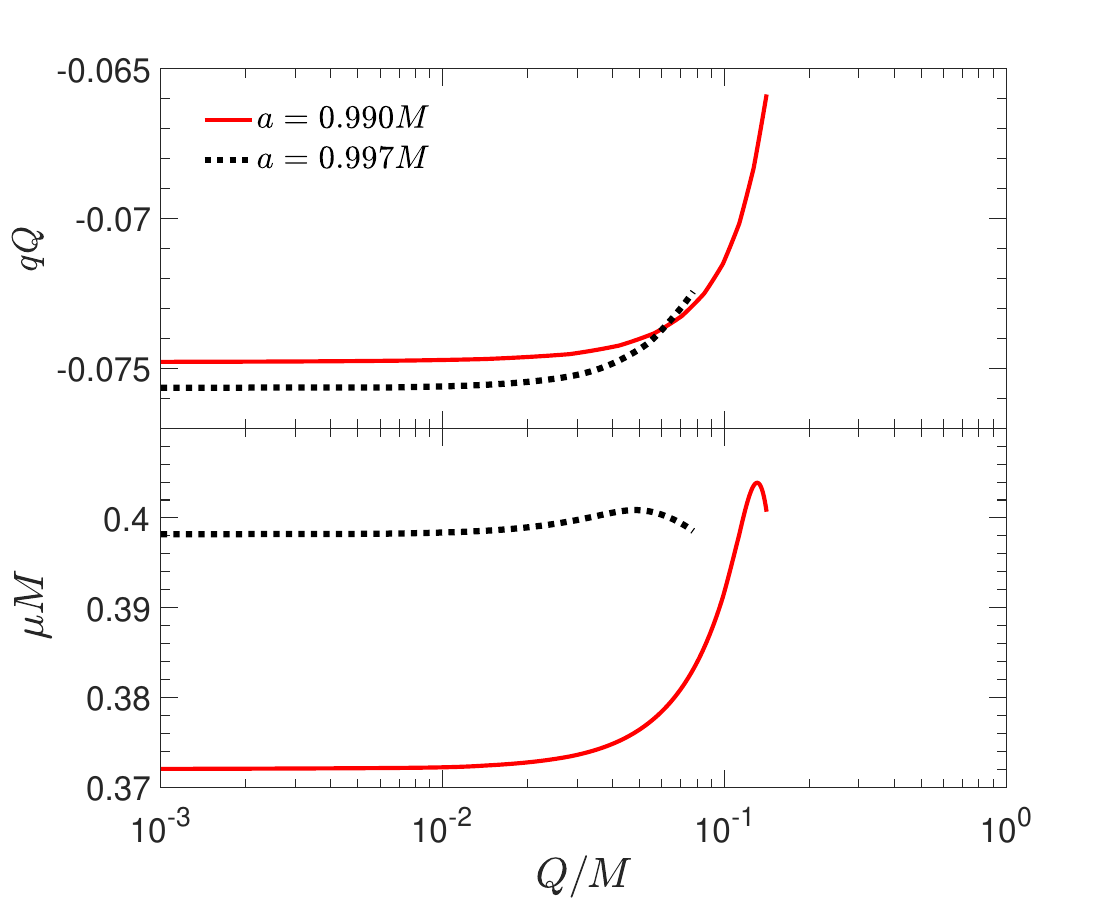}
	\caption{
		The most unstable modes for $l=m=1$.
		Left panel: The imaginary of $\omega_I$ of the most unstable modes as a function of $Q$. 
		Right panel: Dimensionless coupling $qQ$ (top) and $\mu M$ (bottom) of the most unstable modes as a function of $Q$.
		}
	\label{fig: peak_value}
\end{figure*}

Let us now consider the maximum growth rate of the unstable modes. 
From Fig.~\ref{fig: ContourPlot} and the right panel of Fig.~\ref{fig: 3D plot}, we see that for a given $m$ and fixed background parameters $\left\lbrace a,Q\right\rbrace $, 
$\omega_I$ has a peak value inside the unstable region. 
We have computed the bound state frequencies for a wide range of the black hole parameters.
As in the Kerr case, the instability turns out to be most significant for $l=m=1$ and $0.99\leq a/M\leq0.999$.
For each pair of $\left\lbrace a,Q\right\rbrace $, we search the peak value of $\omega_I$ and the corresponding values of $qQ$ and $\mu M$.
It is found that when $a\gtrsim0.997M$, the peak value of $\omega_I$ is decreased as $Q$ grows.
However, when $a\lesssim0.997M$, the peak value of $\omega_I$ does not vary monotonously with $Q$.
The typical results are illustrated in Fig.~\ref{fig: peak_value}, which shows the peak value of $\omega_I$ (left panel)
and the corresponding $qQ$ and $\mu M$ (right panel) as a function of $Q$ for different values of $a$.
The plot shows that when $a=0.99M$, the peak value of $\omega_I$ is greatest at $Q=0.1105M$, 
with $\omega_I M=1.736\times10^{-7}$, $qQ=-0.07$ and $\mu M=0.397$. 
As mentioned before, the peak value of $\omega_I$ decreases as $Q$ grows for $a=0.997M$. 
However, this does not mean that the greatest instability is achieved by setting $Q=0$. From the right panel of Fig.~\ref{fig: peak_value},
the coupling $qQ$ tends to a constant when $Q\rightarrow0$. If $Q$ is too small, the $q$ would be very large. This is illustrated quantitatively in Table.~\ref{table:1}.
For this reason, the largest imaginary part of the frequency we can quote is $\omega_I M=1.788\times10^{-7}$, for $a=0.997M$, $0<Q/M\lesssim10^{-3}$, $qQ=-0.0757$, and $\mu M=0.398$.

\begin{center}
	\begin{table*}
		\caption{Several examples of the most unstable modes in the limit $Q/M\ll1$ for $a=0.997M$ and $l=m=1$.}\label{table:1}
		\begin{tabular}{ |c |c |c |c |c |c |c | }
			\hline
			$Q/M$ & $0.00001$   & $0.0001$    & $0.001$     & $0.002$     & $0.004$     & $0.006$    \\\hline
			$-qQ$ & $0.075668$  & $0.075667$  & $0.075655$  & $0.075648$  & $0.075646$  & $0.075643$ \\\hline
			$\mu M$ & $0.398160$  & $0.398163$  & $0.398186$  & $0.398202$  & $0.39822$   & $0.398247$ \\\hline
			$\omega_IM\;(10^{-7})$ & $1.788169$  & $1.788169$  & $1.788159$  & $1.788131$  & $1.788016$  & $1.787822$ \\
			\hline
		\end{tabular}
	\end{table*}
\end{center}

\subsection{Comparison with analytic results}

In the small coupling parameters limit, i.e., $qQ\ll1$ and $\mu M\ll1$, the imaginary part of the bound state frequency 
is well approximated by \cite{Furuhashi:2004jk}
\begin{equation}\label{Eq: analytic omegaI}
	\omega_I\approx\mu\frac{\delta\nu}{i}\frac{\epsilon^2}{\tilde{n}^3}\sim\mu\;\epsilon^{2l+3},
\end{equation}
where $\epsilon=\mu M-qQ$, $\tilde{n}=l+n+1$, and the explicit form of $\delta\nu$ is given by Eq.(24) in Ref.~\cite{Furuhashi:2004jk}. 
For large values of $|qQ|$ and $\mu M$, Eq.~(\ref{Eq: analytic omegaI}) become less accurate. 
However, we find that Eq.~(\ref{Eq: analytic omegaI}) can still give an excellent 
estimation for the quasibound state frequency for $\mu M\sim|qQ|\sim1$, provided $\epsilon$ is small enough.

\begin{figure}
	\includegraphics[width=0.45\textwidth,height=0.27\textheight]{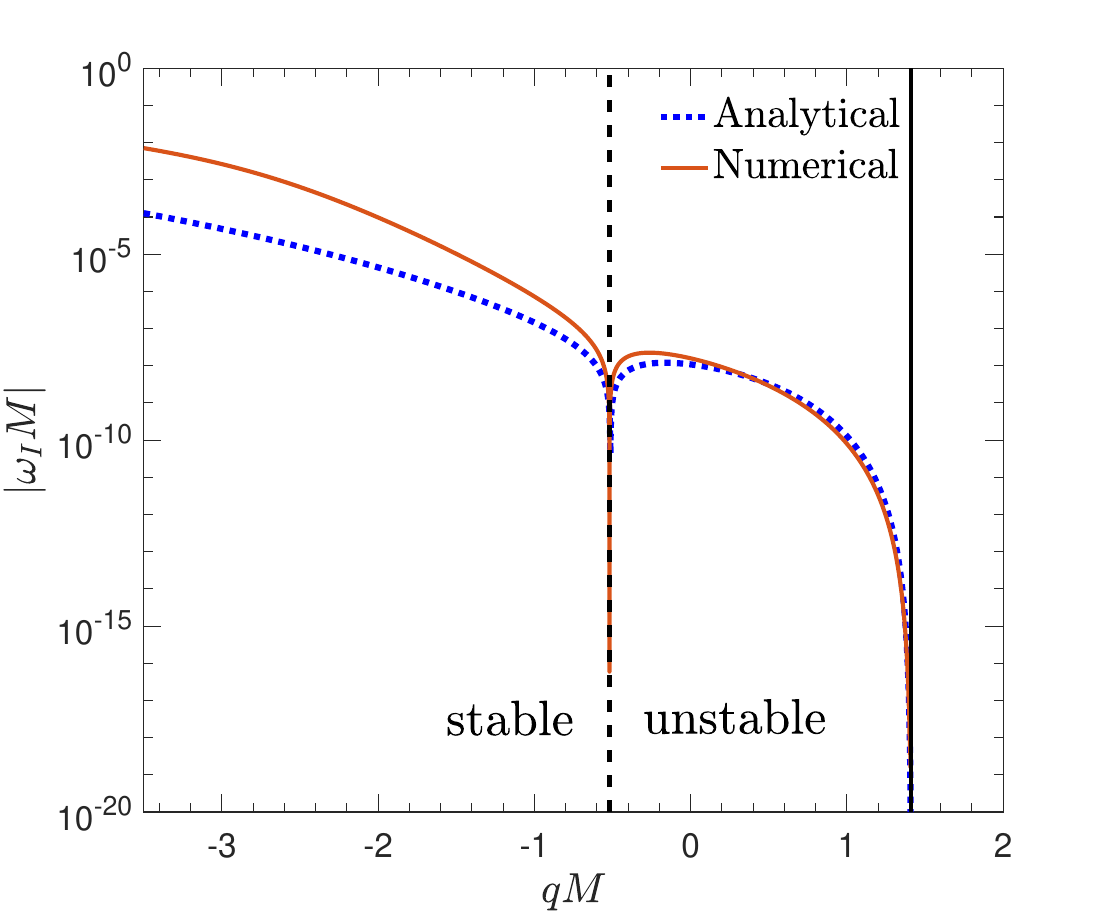}
	\caption{
		Comparison of numerical and analytical results of the imaginary part of the frequency as a function of $qM$ for given $M\mu$ with $a=0.9$, $Q=0.2$, $l=m=1$, and $n=0$. 
		In this plot, the dimensionless mass is fixed as $M\mu=0.282$, which is expect to give the maximum growth rate, as shown in Fig.~\ref{fig: ContourPlot}. 
		The vertical dashed line gives the charge of the stationary cloud, while the vertical solid line corresponds to $qQ=M\mu$.}
	\label{fig: omegaVSq}
\end{figure}

Figure \ref{fig: omegaVSq} compares the numerical and analytical results of $\omega_I$ as a function of $qM$ for $a=0.9M$, $Q=0.2M$, $l=m=1$ and $n=0$.
An overall observation is that the deviation between the numerical and analytical results increases with the absolute value of $qQ$. 
Even inside the unstable region, such deviation is significant: 
the maximum growth rate obtained numerically is about $86\%$ larger than the one obtained from Eq.~(\ref{Eq: analytic omegaI}). 
However, as the charge $q$ increases to approach the threshold condition $qQ=\mu M$, one finds a good agreement between the numerical and analytical results. 
Meanwhile, the scalar field becomes long-lived, since the growth rate tends to zero in this limit. 
Similar observation has been made for charged scalar and Dirac fields around Reissner-Nordstr\"{o}m black holes \cite{Degollado:2013eqa,Sampaio:2014swa,Huang:2017nho}.

\section{Conclusion}\label{Sec: conclusion}

In this work, we have performed a detailed analysis of the superradiant instability of massive charged scalar field in Kerr-Newman spacetime. 
We confirmed the statement in \cite{Huang:2016qnk} that for given values of $\left\lbrace m,a,Q\right\rbrace $, the superradiant instability only
occurs in a finite region in the parameter space of the scalar field.
For the first time, we have computed the bound state spectrum in the whole unstable region,
and found that the maximum mass of the scalar cloud is exactly the upper bound of the mass of the unstable modes, see Figs.~\ref{fig: ContourPlot}-\ref{fig: 3D plot}.

Furthermore, we have also shown that inside the unstable region, there is a peak value of $\omega_I$, which corresponds the most unstable mode
for given values of $\left\lbrace m,a,Q\right\rbrace $. 
We computed the peak values of $\omega_I$ for a range of black hole rotation $0.99\leq a/M\leq0.999$, and the black hole charge $Q/M>10^{-5}$. 
Among the most unstable modes we have computed, we found a maximum value of the imaginary part $\omega_IM=1.788\times10^{-7}$ for $l=m=1$, $\mu M=398$, 
and $qQ=-0.0757$, see Fig.~\ref{fig: peak_value} and Table.~\ref{table:1}. 
As expected, the superradiant instability can be increased by introducing the electromagnetic interaction between the scalar field and black hole.
For example, when $a=0.99M$, Dolan found a maximum growth rate $\tau^{-1}=1.5\times10^{-7}M^{-1}$ of the $l=m=1$ state at $\mu M=0.42$ for a massive scalar field in Kerr spacetime \cite{Dolan:2007mj}. 
For the same rotation parameter $a$, we find that the most unstable mode occurs for $Q=0.1105M$, $qQ=-0.07$ and $\mu M=0.397$, with $\tau^{-1}=1.736\times10^{-7}M^{-1}$.
This is about $15.7\%$ larger than the maximum growth rate in the Kerr case. 

Finally, what will be the final state of the instability when nonlinear effects are taken into account? 
In Kerr case, it turns out to be a rotating black hole embedded in a massive bosonic field that orbiting around the black hole, 
with the frequency precisely matches the horizon angular velocity of the black hole \cite{Herdeiro:2014goa,Herdeiro:2016tmi,East:2017ovw}. 
Hence, we expect that the end state of the superradiant instability discussed above is a Kerr-Newman black hole with scalar hair \cite{Delgado:2016jxq}. 
Perhaps a fully nonlinear analysis is required to confirm this possibility.

\begin{acknowledgments}
This work is supported in part by National Natural Science Foundation of China under Grant No. 11275128, Science and Technology Commission of Shanghai Municipality under Grant No. 12ZR1421700.
\end{acknowledgments}

\bibliography{Ref}

\begin{thebibliography}{38}%
\makeatletter
\providecommand \@ifxundefined [1]{%
 \@ifx{#1\undefined}
}%
\providecommand \@ifnum [1]{%
 \ifnum #1\expandafter \@firstoftwo
 \else \expandafter \@secondoftwo
 \fi
}%
\providecommand \@ifx [1]{%
 \ifx #1\expandafter \@firstoftwo
 \else \expandafter \@secondoftwo
 \fi
}%
\providecommand \natexlab [1]{#1}%
\providecommand \enquote  [1]{``#1''}%
\providecommand \bibnamefont  [1]{#1}%
\providecommand \bibfnamefont [1]{#1}%
\providecommand \citenamefont [1]{#1}%
\providecommand \href@noop [0]{\@secondoftwo}%
\providecommand \href [0]{\begingroup \@sanitize@url \@href}%
\providecommand \@href[1]{\@@startlink{#1}\@@href}%
\providecommand \@@href[1]{\endgroup#1\@@endlink}%
\providecommand \@sanitize@url [0]{\catcode `\\12\catcode `\$12\catcode
  `\&12\catcode `\#12\catcode `\^12\catcode `\_12\catcode `\%12\relax}%
\providecommand \@@startlink[1]{}%
\providecommand \@@endlink[0]{}%
\providecommand \url  [0]{\begingroup\@sanitize@url \@url }%
\providecommand \@url [1]{\endgroup\@href {#1}{\urlprefix }}%
\providecommand \urlprefix  [0]{URL }%
\providecommand \Eprint [0]{\href }%
\providecommand \doibase [0]{http://dx.doi.org/}%
\providecommand \selectlanguage [0]{\@gobble}%
\providecommand \bibinfo  [0]{\@secondoftwo}%
\providecommand \bibfield  [0]{\@secondoftwo}%
\providecommand \translation [1]{[#1]}%
\providecommand \BibitemOpen [0]{}%
\providecommand \bibitemStop [0]{}%
\providecommand \bibitemNoStop [0]{.\EOS\space}%
\providecommand \EOS [0]{\spacefactor3000\relax}%
\providecommand \BibitemShut  [1]{\csname bibitem#1\endcsname}%
\let\auto@bib@innerbib\@empty
\bibitem [{\citenamefont {Regge}\ and\ \citenamefont
  {Wheeler}(1957)}]{Regge:1957td}%
  \BibitemOpen
  \bibfield  {author} {\bibinfo {author} {\bibfnamefont {T.}~\bibnamefont
  {Regge}}\ and\ \bibinfo {author} {\bibfnamefont {J.~A.}\ \bibnamefont
  {Wheeler}},\ }\href {\doibase 10.1103/PhysRev.108.1063} {\bibfield  {journal}
  {\bibinfo  {journal} {Phys. Rev.}\ }\textbf {\bibinfo {volume} {108}},\
  \bibinfo {pages} {1063} (\bibinfo {year} {1957})}\BibitemShut {NoStop}%
\bibitem [{\citenamefont {Zerilli}(1970{\natexlab{a}})}]{Zerilli:1970se}%
  \BibitemOpen
  \bibfield  {author} {\bibinfo {author} {\bibfnamefont {F.~J.}\ \bibnamefont
  {Zerilli}},\ }\href {\doibase 10.1103/PhysRevLett.24.737} {\bibfield
  {journal} {\bibinfo  {journal} {Phys. Rev. Lett.}\ }\textbf {\bibinfo
  {volume} {24}},\ \bibinfo {pages} {737} (\bibinfo {year}
  {1970}{\natexlab{a}})}\BibitemShut {NoStop}%
\bibitem [{\citenamefont {Vishveshwara}(1970)}]{Vishveshwara:1970cc}%
  \BibitemOpen
  \bibfield  {author} {\bibinfo {author} {\bibfnamefont {C.~V.}\ \bibnamefont
  {Vishveshwara}},\ }\href {\doibase 10.1103/PhysRevD.1.2870} {\bibfield
  {journal} {\bibinfo  {journal} {Phys. Rev.}\ }\textbf {\bibinfo {volume}
  {D1}},\ \bibinfo {pages} {2870} (\bibinfo {year} {1970})}\BibitemShut
  {NoStop}%
\bibitem [{\citenamefont {Zerilli}(1970{\natexlab{b}})}]{Zerilli:1971wd}%
  \BibitemOpen
  \bibfield  {author} {\bibinfo {author} {\bibfnamefont {F.~J.}\ \bibnamefont
  {Zerilli}},\ }\href {\doibase 10.1103/PhysRevD.2.2141} {\bibfield  {journal}
  {\bibinfo  {journal} {Phys. Rev.}\ }\textbf {\bibinfo {volume} {D2}},\
  \bibinfo {pages} {2141} (\bibinfo {year} {1970}{\natexlab{b}})}\BibitemShut
  {NoStop}%
\bibitem [{\citenamefont {Teukolsky}(1972)}]{Teukolsky:1972my}%
  \BibitemOpen
  \bibfield  {author} {\bibinfo {author} {\bibfnamefont {S.~A.}\ \bibnamefont
  {Teukolsky}},\ }\href {\doibase 10.1103/PhysRevLett.29.1114} {\bibfield
  {journal} {\bibinfo  {journal} {Phys. Rev. Lett.}\ }\textbf {\bibinfo
  {volume} {29}},\ \bibinfo {pages} {1114} (\bibinfo {year}
  {1972})}\BibitemShut {NoStop}%
\bibitem [{\citenamefont {Chandrasekhar}(1983)}]{Chandrasekhar1983}%
  \BibitemOpen
  \bibfield  {author} {\bibinfo {author} {\bibfnamefont {S.}~\bibnamefont
  {Chandrasekhar}},\ }\href@noop {} {\emph {\bibinfo {title} {The Mathematical
  Theory of Black Holes (International Series of Monographs on Physics)}}}\
  (\bibinfo  {publisher} {Oxford University Press},\ \bibinfo {year}
  {1983})\BibitemShut {NoStop}%
\bibitem [{\citenamefont {Konoplya}\ and\ \citenamefont
  {Zhidenko}(2011)}]{Konoplya:2011qq}%
  \BibitemOpen
  \bibfield  {author} {\bibinfo {author} {\bibfnamefont {R.~A.}\ \bibnamefont
  {Konoplya}}\ and\ \bibinfo {author} {\bibfnamefont {A.}~\bibnamefont
  {Zhidenko}},\ }\href {\doibase 10.1103/RevModPhys.83.793} {\bibfield
  {journal} {\bibinfo  {journal} {Rev. Mod. Phys.}\ }\textbf {\bibinfo {volume}
  {83}},\ \bibinfo {pages} {793} (\bibinfo {year} {2011})},\ \Eprint
  {http://arxiv.org/abs/1102.4014} {arXiv:1102.4014 [gr-qc]} \BibitemShut
  {NoStop}%
\bibitem [{\citenamefont {Newman}\ \emph {et~al.}(1965)\citenamefont {Newman},
  \citenamefont {Couch}, \citenamefont {Chinnapared}, \citenamefont {Exton},
  \citenamefont {Prakash},\ and\ \citenamefont {Torrence}}]{Newman:1965my}%
  \BibitemOpen
  \bibfield  {author} {\bibinfo {author} {\bibfnamefont {E.~T.}\ \bibnamefont
  {Newman}}, \bibinfo {author} {\bibfnamefont {R.}~\bibnamefont {Couch}},
  \bibinfo {author} {\bibfnamefont {K.}~\bibnamefont {Chinnapared}}, \bibinfo
  {author} {\bibfnamefont {A.}~\bibnamefont {Exton}}, \bibinfo {author}
  {\bibfnamefont {A.}~\bibnamefont {Prakash}}, \ and\ \bibinfo {author}
  {\bibfnamefont {R.}~\bibnamefont {Torrence}},\ }\href {\doibase
  10.1063/1.1704351} {\bibfield  {journal} {\bibinfo  {journal} {J. Math.
  Phys.}\ }\textbf {\bibinfo {volume} {6}},\ \bibinfo {pages} {918} (\bibinfo
  {year} {1965})}\BibitemShut {NoStop}%
\bibitem [{\citenamefont {Zilhão}\ \emph {et~al.}(2014)\citenamefont
  {Zilhão}, \citenamefont {Cardoso}, \citenamefont {Herdeiro}, \citenamefont
  {Lehner},\ and\ \citenamefont {Sperhake}}]{Zilhao:2014wqa}%
  \BibitemOpen
  \bibfield  {author} {\bibinfo {author} {\bibfnamefont {M.}~\bibnamefont
  {Zilhão}}, \bibinfo {author} {\bibfnamefont {V.}~\bibnamefont {Cardoso}},
  \bibinfo {author} {\bibfnamefont {C.}~\bibnamefont {Herdeiro}}, \bibinfo
  {author} {\bibfnamefont {L.}~\bibnamefont {Lehner}}, \ and\ \bibinfo {author}
  {\bibfnamefont {U.}~\bibnamefont {Sperhake}},\ }\href {\doibase
  10.1103/PhysRevD.90.124088} {\bibfield  {journal} {\bibinfo  {journal} {Phys.
  Rev.}\ }\textbf {\bibinfo {volume} {D90}},\ \bibinfo {pages} {124088}
  (\bibinfo {year} {2014})},\ \Eprint {http://arxiv.org/abs/1410.0694}
  {arXiv:1410.0694 [gr-qc]} \BibitemShut {NoStop}%
\bibitem [{\citenamefont {Dias}\ \emph {et~al.}(2015)\citenamefont {Dias},
  \citenamefont {Godazgar},\ and\ \citenamefont {Santos}}]{Dias:2015wqa}%
  \BibitemOpen
  \bibfield  {author} {\bibinfo {author} {\bibfnamefont {O.~J.~C.}\
  \bibnamefont {Dias}}, \bibinfo {author} {\bibfnamefont {M.}~\bibnamefont
  {Godazgar}}, \ and\ \bibinfo {author} {\bibfnamefont {J.~E.}\ \bibnamefont
  {Santos}},\ }\href {\doibase 10.1103/PhysRevLett.114.151101} {\bibfield
  {journal} {\bibinfo  {journal} {Phys. Rev. Lett.}\ }\textbf {\bibinfo
  {volume} {114}},\ \bibinfo {pages} {151101} (\bibinfo {year} {2015})},\
  \Eprint {http://arxiv.org/abs/1501.04625} {arXiv:1501.04625 [gr-qc]}
  \BibitemShut {NoStop}%
\bibitem [{\citenamefont {Press}\ and\ \citenamefont
  {Teukolsky}(1972)}]{Press:1972zz}%
  \BibitemOpen
  \bibfield  {author} {\bibinfo {author} {\bibfnamefont {W.~H.}\ \bibnamefont
  {Press}}\ and\ \bibinfo {author} {\bibfnamefont {S.~A.}\ \bibnamefont
  {Teukolsky}},\ }\href {\doibase 10.1038/238211a0} {\bibfield  {journal}
  {\bibinfo  {journal} {Nature}\ }\textbf {\bibinfo {volume} {238}},\ \bibinfo
  {pages} {211} (\bibinfo {year} {1972})}\BibitemShut {NoStop}%
\bibitem [{\citenamefont {Damour}\ \emph {et~al.}(1976)\citenamefont {Damour},
  \citenamefont {Deruelle},\ and\ \citenamefont {Ruffini}}]{Damour:1976kh}%
  \BibitemOpen
  \bibfield  {author} {\bibinfo {author} {\bibfnamefont {T.}~\bibnamefont
  {Damour}}, \bibinfo {author} {\bibfnamefont {N.}~\bibnamefont {Deruelle}}, \
  and\ \bibinfo {author} {\bibfnamefont {R.}~\bibnamefont {Ruffini}},\ }\href
  {\doibase 10.1007/BF02725534} {\bibfield  {journal} {\bibinfo  {journal}
  {Lett. Nuovo Cim.}\ }\textbf {\bibinfo {volume} {15}},\ \bibinfo {pages}
  {257} (\bibinfo {year} {1976})}\BibitemShut {NoStop}%
\bibitem [{\citenamefont {Bekenstein}(1973)}]{Bekenstein:1973mi}%
  \BibitemOpen
  \bibfield  {author} {\bibinfo {author} {\bibfnamefont {J.~D.}\ \bibnamefont
  {Bekenstein}},\ }\href {\doibase 10.1103/PhysRevD.7.949} {\bibfield
  {journal} {\bibinfo  {journal} {Phys. Rev.}\ }\textbf {\bibinfo {volume}
  {D7}},\ \bibinfo {pages} {949} (\bibinfo {year} {1973})}\BibitemShut
  {NoStop}%
\bibitem [{\citenamefont {Brito}\ \emph {et~al.}(2015)\citenamefont {Brito},
  \citenamefont {Cardoso},\ and\ \citenamefont {Pani}}]{Brito:2015oca}%
  \BibitemOpen
  \bibfield  {author} {\bibinfo {author} {\bibfnamefont {R.}~\bibnamefont
  {Brito}}, \bibinfo {author} {\bibfnamefont {V.}~\bibnamefont {Cardoso}}, \
  and\ \bibinfo {author} {\bibfnamefont {P.}~\bibnamefont {Pani}},\ }\href
  {\doibase 10.1007/978-3-319-19000-6} {\bibfield  {journal} {\bibinfo
  {journal} {Lect. Notes Phys.}\ }\textbf {\bibinfo {volume} {906}},\ \bibinfo
  {pages} {pp.1} (\bibinfo {year} {2015})},\ \Eprint
  {http://arxiv.org/abs/1501.06570} {arXiv:1501.06570 [gr-qc]} \BibitemShut
  {NoStop}%
\bibitem [{\citenamefont {Zouros}\ and\ \citenamefont
  {Eardley}(1979)}]{Zouros:1979iw}%
  \BibitemOpen
  \bibfield  {author} {\bibinfo {author} {\bibfnamefont {T.~J.~M.}\
  \bibnamefont {Zouros}}\ and\ \bibinfo {author} {\bibfnamefont {D.~M.}\
  \bibnamefont {Eardley}},\ }\href {\doibase 10.1016/0003-4916(79)90237-9}
  {\bibfield  {journal} {\bibinfo  {journal} {Annals Phys.}\ }\textbf {\bibinfo
  {volume} {118}},\ \bibinfo {pages} {139} (\bibinfo {year}
  {1979})}\BibitemShut {NoStop}%
\bibitem [{\citenamefont {Detweiler}(1980)}]{Detweiler:1980uk}%
  \BibitemOpen
  \bibfield  {author} {\bibinfo {author} {\bibfnamefont {S.~L.}\ \bibnamefont
  {Detweiler}},\ }\href {\doibase 10.1103/PhysRevD.22.2323} {\bibfield
  {journal} {\bibinfo  {journal} {Phys. Rev.}\ }\textbf {\bibinfo {volume}
  {D22}},\ \bibinfo {pages} {2323} (\bibinfo {year} {1980})}\BibitemShut
  {NoStop}%
\bibitem [{\citenamefont {Furuhashi}\ and\ \citenamefont
  {Nambu}(2004)}]{Furuhashi:2004jk}%
  \BibitemOpen
  \bibfield  {author} {\bibinfo {author} {\bibfnamefont {H.}~\bibnamefont
  {Furuhashi}}\ and\ \bibinfo {author} {\bibfnamefont {Y.}~\bibnamefont
  {Nambu}},\ }\href {\doibase 10.1143/PTP.112.983} {\bibfield  {journal}
  {\bibinfo  {journal} {Prog. Theor. Phys.}\ }\textbf {\bibinfo {volume}
  {112}},\ \bibinfo {pages} {983} (\bibinfo {year} {2004})},\ \Eprint
  {http://arxiv.org/abs/gr-qc/0402037} {arXiv:gr-qc/0402037 [gr-qc]}
  \BibitemShut {NoStop}%
\bibitem [{\citenamefont {Dolan}(2007)}]{Dolan:2007mj}%
  \BibitemOpen
  \bibfield  {author} {\bibinfo {author} {\bibfnamefont {S.~R.}\ \bibnamefont
  {Dolan}},\ }\href {\doibase 10.1103/PhysRevD.76.084001} {\bibfield  {journal}
  {\bibinfo  {journal} {Phys. Rev.}\ }\textbf {\bibinfo {volume} {D76}},\
  \bibinfo {pages} {084001} (\bibinfo {year} {2007})},\ \Eprint
  {http://arxiv.org/abs/0705.2880} {arXiv:0705.2880 [gr-qc]} \BibitemShut
  {NoStop}%
\bibitem [{\citenamefont {Rosa}(2010)}]{Rosa:2009ei}%
  \BibitemOpen
  \bibfield  {author} {\bibinfo {author} {\bibfnamefont {J.~G.}\ \bibnamefont
  {Rosa}},\ }\href {\doibase 10.1007/JHEP06(2010)015} {\bibfield  {journal}
  {\bibinfo  {journal} {JHEP}\ }\textbf {\bibinfo {volume} {06}},\ \bibinfo
  {pages} {015} (\bibinfo {year} {2010})},\ \Eprint
  {http://arxiv.org/abs/0912.1780} {arXiv:0912.1780 [hep-th]} \BibitemShut
  {NoStop}%
\bibitem [{\citenamefont {Hod}\ and\ \citenamefont {Hod}(2010)}]{Hod:2009cp}%
  \BibitemOpen
  \bibfield  {author} {\bibinfo {author} {\bibfnamefont {S.}~\bibnamefont
  {Hod}}\ and\ \bibinfo {author} {\bibfnamefont {O.}~\bibnamefont {Hod}},\
  }\href {\doibase 10.1103/PhysRevD.81.061502} {\bibfield  {journal} {\bibinfo
  {journal} {Phys. Rev.}\ }\textbf {\bibinfo {volume} {D81}},\ \bibinfo {pages}
  {061502} (\bibinfo {year} {2010})},\ \Eprint {http://arxiv.org/abs/0910.0734}
  {arXiv:0910.0734 [gr-qc]} \BibitemShut {NoStop}%
\bibitem [{\citenamefont {Huang}\ and\ \citenamefont
  {Liu}(2016)}]{Huang:2016qnk}%
  \BibitemOpen
  \bibfield  {author} {\bibinfo {author} {\bibfnamefont {Y.}~\bibnamefont
  {Huang}}\ and\ \bibinfo {author} {\bibfnamefont {D.-J.}\ \bibnamefont
  {Liu}},\ }\href {\doibase 10.1103/PhysRevD.94.064030} {\bibfield  {journal}
  {\bibinfo  {journal} {Phys. Rev.}\ }\textbf {\bibinfo {volume} {D94}},\
  \bibinfo {pages} {064030} (\bibinfo {year} {2016})},\ \Eprint
  {http://arxiv.org/abs/1606.08913} {arXiv:1606.08913 [gr-qc]} \BibitemShut
  {NoStop}%
\bibitem [{\citenamefont {Dolan}(2013)}]{Dolan:2012yt}%
  \BibitemOpen
  \bibfield  {author} {\bibinfo {author} {\bibfnamefont {S.~R.}\ \bibnamefont
  {Dolan}},\ }\href {\doibase 10.1103/PhysRevD.87.124026} {\bibfield  {journal}
  {\bibinfo  {journal} {Phys. Rev.}\ }\textbf {\bibinfo {volume} {D87}},\
  \bibinfo {pages} {124026} (\bibinfo {year} {2013})},\ \Eprint
  {http://arxiv.org/abs/1212.1477} {arXiv:1212.1477 [gr-qc]} \BibitemShut
  {NoStop}%
\bibitem [{\citenamefont {Olver}\ \emph {et~al.}(2010)\citenamefont {Olver},
  \citenamefont {Lozier}, \citenamefont {Boisvert},\ and\ \citenamefont
  {Clark}}]{Olver:2010:NHMF}%
  \BibitemOpen
  \bibinfo {editor} {\bibfnamefont {F.~W.~J.}\ \bibnamefont {Olver}}, \bibinfo
  {editor} {\bibfnamefont {D.~W.}\ \bibnamefont {Lozier}}, \bibinfo {editor}
  {\bibfnamefont {R.~F.}\ \bibnamefont {Boisvert}}, \ and\ \bibinfo {editor}
  {\bibfnamefont {C.~W.}\ \bibnamefont {Clark}},\ eds.,\ \href@noop {} {\emph
  {\bibinfo {title} {{NIST Handbook of Mathematical Functions}}}}\ (\bibinfo
  {publisher} {Cambridge University Press},\ \bibinfo {address} {New York,
  NY},\ \bibinfo {year} {2010})\ \bibinfo {note} {print companion to
  \cite{NIST:DLMF}}\BibitemShut {NoStop}%
\bibitem [{{\relax DLMF}()}]{NIST:DLMF}%
  \BibitemOpen
  {\relax DLMF},\ \href {http://dlmf.nist.gov/} {\enquote {\bibinfo {title}
  {{NIST Digital Library of Mathematical Functions}},}\ }\bibinfo
  {howpublished} {http://dlmf.nist.gov/, Release 1.0.11 of 2016-06-08},\
  \bibinfo {note} {online companion to \cite{Olver:2010:NHMF}}\BibitemShut
  {NoStop}%
\bibitem [{\citenamefont {Leaver}(1985)}]{Leaver:1985ax}%
  \BibitemOpen
  \bibfield  {author} {\bibinfo {author} {\bibfnamefont {E.~W.}\ \bibnamefont
  {Leaver}},\ }\href {\doibase 10.1098/rspa.1985.0119} {\bibfield  {journal}
  {\bibinfo  {journal} {Proc. Roy. Soc. Lond.}\ }\textbf {\bibinfo {volume}
  {A402}},\ \bibinfo {pages} {285} (\bibinfo {year} {1985})}\BibitemShut
  {NoStop}%
\bibitem [{\citenamefont {Benone}\ \emph {et~al.}(2014)\citenamefont {Benone},
  \citenamefont {Crispino}, \citenamefont {Herdeiro},\ and\ \citenamefont
  {Radu}}]{Benone:2014ssa}%
  \BibitemOpen
  \bibfield  {author} {\bibinfo {author} {\bibfnamefont {C.~L.}\ \bibnamefont
  {Benone}}, \bibinfo {author} {\bibfnamefont {L.~C.~B.}\ \bibnamefont
  {Crispino}}, \bibinfo {author} {\bibfnamefont {C.}~\bibnamefont {Herdeiro}},
  \ and\ \bibinfo {author} {\bibfnamefont {E.}~\bibnamefont {Radu}},\ }\href
  {\doibase 10.1103/PhysRevD.90.104024} {\bibfield  {journal} {\bibinfo
  {journal} {Phys. Rev.}\ }\textbf {\bibinfo {volume} {D90}},\ \bibinfo {pages}
  {104024} (\bibinfo {year} {2014})},\ \Eprint {http://arxiv.org/abs/1409.1593}
  {arXiv:1409.1593 [gr-qc]} \BibitemShut {NoStop}%
\bibitem [{\citenamefont {Hod}(2012)}]{Hod:2012px}%
  \BibitemOpen
  \bibfield  {author} {\bibinfo {author} {\bibfnamefont {S.}~\bibnamefont
  {Hod}},\ }\href {\doibase 10.1103/PhysRevD.86.129902,
  10.1103/PhysRevD.86.104026} {\bibfield  {journal} {\bibinfo  {journal} {Phys.
  Rev.}\ }\textbf {\bibinfo {volume} {D86}},\ \bibinfo {pages} {104026}
  (\bibinfo {year} {2012})},\ \bibinfo {note} {[Erratum: Phys.
  Rev.D86,129902(2012)]},\ \Eprint {http://arxiv.org/abs/1211.3202}
  {arXiv:1211.3202 [gr-qc]} \BibitemShut {NoStop}%
\bibitem [{\citenamefont {Herdeiro}\ and\ \citenamefont
  {Radu}(2014)}]{Herdeiro:2014goa}%
  \BibitemOpen
  \bibfield  {author} {\bibinfo {author} {\bibfnamefont {C.~A.~R.}\
  \bibnamefont {Herdeiro}}\ and\ \bibinfo {author} {\bibfnamefont
  {E.}~\bibnamefont {Radu}},\ }\href {\doibase 10.1103/PhysRevLett.112.221101}
  {\bibfield  {journal} {\bibinfo  {journal} {Phys. Rev. Lett.}\ }\textbf
  {\bibinfo {volume} {112}},\ \bibinfo {pages} {221101} (\bibinfo {year}
  {2014})},\ \Eprint {http://arxiv.org/abs/1403.2757} {arXiv:1403.2757 [gr-qc]}
  \BibitemShut {NoStop}%
\bibitem [{\citenamefont {Hod}(2015)}]{Hod:2016yxg}%
  \BibitemOpen
  \bibfield  {author} {\bibinfo {author} {\bibfnamefont {S.}~\bibnamefont
  {Hod}},\ }\href {\doibase 10.1088/0264-9381/32/13/134002} {\bibfield
  {journal} {\bibinfo  {journal} {Class. Quant. Grav.}\ }\textbf {\bibinfo
  {volume} {32}},\ \bibinfo {pages} {134002} (\bibinfo {year} {2015})},\
  \Eprint {http://arxiv.org/abs/1607.00003} {arXiv:1607.00003 [gr-qc]}
  \BibitemShut {NoStop}%
\bibitem [{\citenamefont {Hod}(2014)}]{Hod:2014baa}%
  \BibitemOpen
  \bibfield  {author} {\bibinfo {author} {\bibfnamefont {S.}~\bibnamefont
  {Hod}},\ }\href {\doibase 10.1103/PhysRevD.90.024051} {\bibfield  {journal}
  {\bibinfo  {journal} {Phys. Rev.}\ }\textbf {\bibinfo {volume} {D90}},\
  \bibinfo {pages} {024051} (\bibinfo {year} {2014})},\ \Eprint
  {http://arxiv.org/abs/1406.1179} {arXiv:1406.1179 [gr-qc]} \BibitemShut
  {NoStop}%
\bibitem [{\citenamefont {Bernard}(2016)}]{Bernard:2016wqo}%
  \BibitemOpen
  \bibfield  {author} {\bibinfo {author} {\bibfnamefont {C.}~\bibnamefont
  {Bernard}},\ }\href {\doibase 10.1103/PhysRevD.94.085007} {\bibfield
  {journal} {\bibinfo  {journal} {Phys. Rev.}\ }\textbf {\bibinfo {volume}
  {D94}},\ \bibinfo {pages} {085007} (\bibinfo {year} {2016})},\ \Eprint
  {http://arxiv.org/abs/1608.05974} {arXiv:1608.05974 [gr-qc]} \BibitemShut
  {NoStop}%
\bibitem [{\citenamefont {Huang}\ \emph
  {et~al.}(2017{\natexlab{a}})\citenamefont {Huang}, \citenamefont {Liu},
  \citenamefont {Zhai},\ and\ \citenamefont {Li}}]{Huang:2017whw}%
  \BibitemOpen
  \bibfield  {author} {\bibinfo {author} {\bibfnamefont {Y.}~\bibnamefont
  {Huang}}, \bibinfo {author} {\bibfnamefont {D.-J.}\ \bibnamefont {Liu}},
  \bibinfo {author} {\bibfnamefont {X.-H.}\ \bibnamefont {Zhai}}, \ and\
  \bibinfo {author} {\bibfnamefont {X.-Z.}\ \bibnamefont {Li}},\ }\href
  {\doibase 10.1088/1361-6382/aa7964} {\bibfield  {journal} {\bibinfo
  {journal} {Class. Quant. Grav.}\ }\textbf {\bibinfo {volume} {34}},\ \bibinfo
  {pages} {155002} (\bibinfo {year} {2017}{\natexlab{a}})},\ \Eprint
  {http://arxiv.org/abs/1706.04441} {arXiv:1706.04441 [gr-qc]} \BibitemShut
  {NoStop}%
\bibitem [{\citenamefont {Degollado}\ and\ \citenamefont
  {Herdeiro}(2013)}]{Degollado:2013eqa}%
  \BibitemOpen
  \bibfield  {author} {\bibinfo {author} {\bibfnamefont {J.~C.}\ \bibnamefont
  {Degollado}}\ and\ \bibinfo {author} {\bibfnamefont {C.~A.~R.}\ \bibnamefont
  {Herdeiro}},\ }\href {\doibase 10.1007/s10714-013-1598-6} {\bibfield
  {journal} {\bibinfo  {journal} {Gen. Rel. Grav.}\ }\textbf {\bibinfo {volume}
  {45}},\ \bibinfo {pages} {2483} (\bibinfo {year} {2013})},\ \Eprint
  {http://arxiv.org/abs/1303.2392} {arXiv:1303.2392 [gr-qc]} \BibitemShut
  {NoStop}%
\bibitem [{\citenamefont {Sampaio}\ \emph {et~al.}(2014)\citenamefont
  {Sampaio}, \citenamefont {Herdeiro},\ and\ \citenamefont
  {Wang}}]{Sampaio:2014swa}%
  \BibitemOpen
  \bibfield  {author} {\bibinfo {author} {\bibfnamefont {M.~O.~P.}\
  \bibnamefont {Sampaio}}, \bibinfo {author} {\bibfnamefont {C.}~\bibnamefont
  {Herdeiro}}, \ and\ \bibinfo {author} {\bibfnamefont {M.}~\bibnamefont
  {Wang}},\ }\href {\doibase 10.1103/PhysRevD.90.064004} {\bibfield  {journal}
  {\bibinfo  {journal} {Phys. Rev.}\ }\textbf {\bibinfo {volume} {D90}},\
  \bibinfo {pages} {064004} (\bibinfo {year} {2014})},\ \Eprint
  {http://arxiv.org/abs/1406.3536} {arXiv:1406.3536 [gr-qc]} \BibitemShut
  {NoStop}%
\bibitem [{\citenamefont {Huang}\ \emph
  {et~al.}(2017{\natexlab{b}})\citenamefont {Huang}, \citenamefont {Liu},
  \citenamefont {Zhai},\ and\ \citenamefont {Li}}]{Huang:2017nho}%
  \BibitemOpen
  \bibfield  {author} {\bibinfo {author} {\bibfnamefont {Y.}~\bibnamefont
  {Huang}}, \bibinfo {author} {\bibfnamefont {D.-J.}\ \bibnamefont {Liu}},
  \bibinfo {author} {\bibfnamefont {X.-h.}\ \bibnamefont {Zhai}}, \ and\
  \bibinfo {author} {\bibfnamefont {X.-z.}\ \bibnamefont {Li}},\ }\href
  {\doibase 10.1103/PhysRevD.96.065002} {\bibfield  {journal} {\bibinfo
  {journal} {Phys. Rev.}\ }\textbf {\bibinfo {volume} {D96}},\ \bibinfo {pages}
  {065002} (\bibinfo {year} {2017}{\natexlab{b}})},\ \Eprint
  {http://arxiv.org/abs/1708.04761} {arXiv:1708.04761 [gr-qc]} \BibitemShut
  {NoStop}%
\bibitem [{\citenamefont {Herdeiro}\ \emph {et~al.}(2016)\citenamefont
  {Herdeiro}, \citenamefont {Radu},\ and\ \citenamefont
  {Runarsson}}]{Herdeiro:2016tmi}%
  \BibitemOpen
  \bibfield  {author} {\bibinfo {author} {\bibfnamefont {C.}~\bibnamefont
  {Herdeiro}}, \bibinfo {author} {\bibfnamefont {E.}~\bibnamefont {Radu}}, \
  and\ \bibinfo {author} {\bibfnamefont {H.}~\bibnamefont {Runarsson}},\ }\href
  {\doibase 10.1088/0264-9381/33/15/154001} {\bibfield  {journal} {\bibinfo
  {journal} {Class. Quant. Grav.}\ }\textbf {\bibinfo {volume} {33}},\ \bibinfo
  {pages} {154001} (\bibinfo {year} {2016})},\ \Eprint
  {http://arxiv.org/abs/1603.02687} {arXiv:1603.02687 [gr-qc]} \BibitemShut
  {NoStop}%
\bibitem [{\citenamefont {East}\ and\ \citenamefont
  {Pretorius}(2017)}]{East:2017ovw}%
  \BibitemOpen
  \bibfield  {author} {\bibinfo {author} {\bibfnamefont {W.~E.}\ \bibnamefont
  {East}}\ and\ \bibinfo {author} {\bibfnamefont {F.}~\bibnamefont
  {Pretorius}},\ }\href {\doibase 10.1103/PhysRevLett.119.041101} {\bibfield
  {journal} {\bibinfo  {journal} {Phys. Rev. Lett.}\ }\textbf {\bibinfo
  {volume} {119}},\ \bibinfo {pages} {041101} (\bibinfo {year} {2017})},\
  \Eprint {http://arxiv.org/abs/1704.04791} {arXiv:1704.04791 [gr-qc]}
  \BibitemShut {NoStop}%
\bibitem [{\citenamefont {Delgado}\ \emph {et~al.}(2016)\citenamefont
  {Delgado}, \citenamefont {Herdeiro}, \citenamefont {Radu},\ and\
  \citenamefont {Runarsson}}]{Delgado:2016jxq}%
  \BibitemOpen
  \bibfield  {author} {\bibinfo {author} {\bibfnamefont {J.~F.~M.}\
  \bibnamefont {Delgado}}, \bibinfo {author} {\bibfnamefont {C.~A.~R.}\
  \bibnamefont {Herdeiro}}, \bibinfo {author} {\bibfnamefont {E.}~\bibnamefont
  {Radu}}, \ and\ \bibinfo {author} {\bibfnamefont {H.}~\bibnamefont
  {Runarsson}},\ }\href {\doibase 10.1016/j.physletb.2016.08.032} {\bibfield
  {journal} {\bibinfo  {journal} {Phys. Lett.}\ }\textbf {\bibinfo {volume}
  {B761}},\ \bibinfo {pages} {234} (\bibinfo {year} {2016})},\ \Eprint
  {http://arxiv.org/abs/1608.00631} {arXiv:1608.00631 [gr-qc]} \BibitemShut
  {NoStop}%
\end{thebibliography}%

\end{document}